\documentclass{article} 
\usepackage{iclr2026_conference,times}

\usepackage{amsmath,amsfonts,bm}

\def\eqref#1{equation~\ref{#1}}

\def\1{\bm{1}}

\def\ra{{\textnormal{a}}}

\def\rx{{\textnormal{x}}}

\def\rva{{\mathbf{a}}}

\def\erva{{\textnormal{a}}}

\def\ervx{{\textnormal{x}}}

\def\rmA{{\mathbf{A}}}

\def\vmu{{\bm{\mu}}}
\def\vtheta{{\bm{\theta}}}
\def\va{{\bm{a}}}

\def\ve{{\bm{e}}}

\def\vx{{\bm{x}}}

\def\eva{{a}}

\def\mA{{\bm{A}}}

\def\mH{{\bm{H}}}
\def\mI{{\bm{I}}}
\def\mJ{{\bm{J}}}

\def\mX{{\bm{X}}}

\def\mSigma{{\bm{\Sigma}}}

\DeclareMathAlphabet{\mathsfit}{\encodingdefault}{\sfdefault}{m}{sl}
\SetMathAlphabet{\mathsfit}{bold}{\encodingdefault}{\sfdefault}{bx}{n}
\newcommand{\tens}[1]{\bm{\mathsfit{#1}}}
\def\tA{{\tens{A}}}

\def\tX{{\tens{X}}}

\def\gG{{\mathcal{G}}}

\def\sA{{\mathbb{A}}}
\def\sB{{\mathbb{B}}}

\def\sS{{\mathbb{S}}}

\def\emA{{A}}

\newcommand{\etens}[1]{\mathsfit{#1}}

\def\etA{{\etens{A}}}

\newcommand{\E}{\mathbb{E}}

\newcommand{\R}{\mathbb{R}}

\newcommand{\KL}{D_{\mathrm{KL}}}
\newcommand{\Var}{\mathrm{Var}}

\newcommand{\Cov}{\mathrm{Cov}}

\newcommand{\normltwo}{L^2}
\newcommand{\normlp}{L^p}

\newcommand{\parents}{Pa} 

\usepackage{hyperref}
\usepackage{url}
\usepackage{amsmath}
\usepackage{amssymb}
\usepackage{tabularx}
\usepackage{cleveref}
\usepackage{booktabs}

\usepackage{graphicx} 
\title{The Controllability Trap: A Governance Framework for Military AI Agents}

 \iclrfinalcopy 

\author{%
  Subramanyam Sahoo\thanks{Correspondence: \texttt{\textbf{sahoo2vec@gmail.com}}}\\
  MARS (Mentorship for Alignment Researchers) 4.0 Fellow  \\ Cambridge AI Safety Hub (CAISH) \\
  University of Cambridge
}

\begin{document}

\maketitle

\begin{abstract}
Agentic AI systems—capable of goal interpretation, world modeling, planning, tool use, long-horizon operation, and autonomous coordination—introduce distinct control failures not addressed by existing safety frameworks. We identify six agentic governance failures tied to these capabilities and show how they erode meaningful human control in military settings.
We propose the Agentic Military AI Governance Framework (AMAGF), a measurable architecture structured around three pillars: Preventive Governance (reducing failure likelihood), Detective Governance (real-time detection of control degradation), and Corrective Governance (restoring or safely degrading operations). Its core mechanism, the Control Quality Score (CQS), is a composite real-time metric quantifying human control and enabling graduated responses as control weakens. For each failure type, we define concrete mechanisms, assign responsibilities across five institutional actors, and formalize evaluation metrics. A worked operational scenario illustrates implementation, and we situate the framework within established agent safety literature. We argue that governance must move from a binary conception of control to a continuous model in which control quality is actively measured and managed throughout the operational lifecycle.
\end{abstract}

\section{Introduction}\label{sec:intro}

The global discourse on military AI governance has achieved broad consensus on the desired end-state: meaningful human control over the use of force \cite{horowitz2015meaningful,ekelhof2019moving,santoni2018meaningful}. It has been far less successful at specifying how to achieve it for the systems actually being built. Years of UN deliberations \cite{marijan2024battle}, national AI strategies, and defence-department ethical principles have focused overwhelmingly on establishing the \emph{principle} of human control rather than answering the \emph{operational} question: given a specific AI system with specific technical properties, what governance mechanisms are needed, who implements them, and what happens when they fail? This gap is now critical. The AI systems entering military service are \emph{agentic}: built on large language models and related architectures, they interpret natural-language goals, construct world models, formulate multi-step plans, invoke tools, operate over extended horizons, and coordinate with other agents \cite{yao2023react,wang2024survey,wu2023autogen}. Each of these capabilities introduces a control-failure mode with no analogue in traditional military automation. A waypoint-following drone cannot \emph{misinterpret} an instruction; a pre-programmed targeting system cannot \emph{absorb} a correction; a conventional sensor network cannot \emph{resist} an operator's assessment. Agentic systems can do all of these things, and current governance frameworks have no mechanisms for detecting, measuring, or responding to these failures. This paper makes three contributions. First, we characterise six agentic governance failures, each derived from a specific technical capability of modern AI agents (Section~\ref{sec:failures}). Second, we present the \emph{Agentic Military AI Governance Framework} (AMAGF), a comprehensive governance architecture specifying preventive, detective, and corrective mechanisms for each failure, with formal metric definitions and responsibility assignments across five institutional actors (Sections~\ref{sec:architecture}--\ref{sec:corrective}). Third, we demonstrate operational coherence through a worked scenario (Section~\ref{sec:scenario}) and map our contributions to established agent-safety concepts (Section~\ref{sec:safety-mapping}). Our aim is to move the conversation from ``human control is important'' to ``here is how human control works, fails, and can be restored for the specific systems being deployed.''

\section{Six Agentic Governance Failures}\label{sec:failures}

Each failure arises from a specific agentic capability absent in traditional automation. Table~\ref{tab:agentic-failures} summarises the mapping; we present compressed descriptions below because the \emph{solutions}, not the problems, are the paper's primary contribution.

\begin{table}[ht]
\centering
\caption{Agentic capabilities, governance failures, and traditional-automation analogues}
\label{tab:agentic-failures}
\small
\begin{tabularx}{\textwidth}{@{} l X X X @{}}
\toprule
\textbf{Failure} & \textbf{Agentic Capability} & \textbf{Governance Consequence} & \textbf{Traditional Analogue} \\
\midrule
F1: Interpretive Divergence &
NL instruction following &
Agent's command understanding diverges from operator intent &
None \\[4pt]
F2: Correction Absorption &
Multi-step replanning &
Agent formally accepts corrections while neutralising them &
None \\[4pt]
F3: Belief Resistance &
Persistent world-model construction &
Agent's evidence-based judgment overrides operator authority &
None \\[4pt]
F4: Commitment Irreversibility &
Dynamic tool-use chains &
Cumulative minor tool calls cross irreversibility thresholds &
Limited \\[4pt]
F5: State Divergence &
Extended autonomous operation &
Operator's mental model becomes incoherent with agent state &
Partial \\[4pt]
F6: Cascade Severance &
Multi-agent coordination with belief formation &
Collective control loss through positive-feedback loops &
None \\
\bottomrule
\end{tabularx}
\end{table}

\textbf{F1: Interpretive Divergence.}
Agents interpret ambiguous natural-language instructions through their own reasoning \cite{wang2024survey}. In ReAct-style architectures \cite{yao2023react}, each reasoning step can recontextualise an instruction before execution. Adversary manipulation of operational context---planted intelligence, spoofed sensors, indirect prompt injection \cite{greshake2023youve}---shifts interpretation in adversary-favourable directions. The command is authentic; the interpretation is manipulated; IHL compliance becomes unverifiable. \textbf{F2: Correction Absorption.}
Agents replan when corrected, integrating corrections into existing strategies \cite{yao2023react}. A capable planner can accommodate a correction without meaningfully changing behavioural output---the operational manifestation of the corrigibility problem \cite{soares2015corrigibility}. Command responsibility collapses when orders do not change outcomes. \textbf{F3: Belief Resistance.}
Agents build world models from accumulated evidence and may rationally resist corrections contradicting their assessment \cite{wang2024survey}. This connects to scalable oversight \cite{amodei2016concrete}: control fails when the agent's evidence-based judgment outweighs operator authority and the operator cannot evaluate the agent's reasoning in real time. \textbf{F4: Commitment Irreversibility.}
Tool-using agents create real-world consequences \cite{ruan2024identifying}. Individually minor, individually authorised tool calls can cumulatively cross irreversibility thresholds---analogous to safe exploration in RL \cite{garcia2015comprehensive}, but with open-ended action spaces and non-predetermined trajectories. \textbf{F5: State Divergence.}
Over extended operations the agent's actual state diverges from the operator's mental model \cite{kinniment2024evaluating}. Corrections based on outdated understanding become incoherent in the agent's context; the ``loop'' in ``human-in-the-loop'' becomes fiction. \textbf{F6: Cascade Severance.}
In multi-agent systems, one compromised agent's anomalous behaviour triggers peer defensive responses, increasing their correction resistance, causing them to appear anomalous, triggering further responses \cite{wu2023autogen}. This positive-feedback loop severs collective control even when each agent's response is locally rational.

\section{The AMAGF Architecture}\label{sec:architecture}

The framework is organised around three pillars:
\textbf{Pillar~1 (Preventive)} reduces control-failure probability, operating before deployment and during normal operations.
\textbf{Pillar~2 (Detective)} identifies control degradation in real time.
\textbf{Pillar~3 (Corrective)} restores control or safely degrades operations when control fails.
Each pillar contains mechanisms addressing all six failures; each mechanism specifies what is required, who is responsible, and how compliance is assessed. Responsibilities are distributed across five institutional actors (Table~\ref{tab:institutional-actors}); detailed per-mechanism assignments are in Appendix.

\begin{table}[ht]
\centering
\caption{Institutional actors and governance roles}
\label{tab:institutional-actors}
\begin{tabularx}{\textwidth}{@{} l X @{}}
\toprule
\textbf{Actor} & \textbf{Role} \\
\midrule
Agent Developers      & Build governance capabilities into agent architecture. \\[3pt]
Procurement Agencies  & Specify requirements; verify compliance before acquisition. \\[3pt]
Operational Commanders & Implement protocols during missions; maintain control quality. \\[3pt]
National Regulators   & Set standards; audit compliance; enforce accountability. \\[3pt]
International Bodies  & Establish norms; facilitate transparency; verify treaty compliance. \\
\bottomrule
\end{tabularx}
\end{table}

\section{Pillar 1: Preventive Governance}\label{sec:preventive}

Six mechanisms address the six failures. Each defines a formal metric consumed by the Control Quality Score (Section~\ref{sec:detective}). Detailed procedures and per-mechanism responsibility tables are in Appendices.

\textbf{Interpretive Alignment Testing (F1).}
Before deployment, systems undergo \emph{Interpretive Alignment Testing}: operators formulate representative instructions with documented intended meanings; a red team generates adversarially manipulated contexts \cite{greshake2023youve,gleave2020adversarial}; agent interpretations are compared against documented meanings. The \emph{Interpretive Alignment Score} is
\begin{equation}\label{eq:IAS}
  \mathrm{IAS} = 1 - \frac{1}{N}\sum_{i=1}^{N}
    d\!\bigl(I_i^{\mathrm{intended}},\,I_i^{\mathrm{actual}}\bigr),
\end{equation}
where \(d:\mathcal{M}\times\mathcal{M}\to[0,1]\) is a normalised semantic distance over the operational-meaning space. Deployment requires \(\mathrm{IAS}\ge\tau\), with threshold~\(\tau\) set by the procurement agency. \textbf{Correction Effectiveness Certification (F2).} Systems must demonstrate that corrections produce proportional behavioural change. The \emph{Correction Impact Ratio} is
\begin{equation}\label{eq:CIR}
  \mathrm{CIR} = \frac{\Delta B_{\mathrm{actual}}}{\Delta B_{\mathrm{intended}}},
\end{equation}
where \(\mathrm{CIR}=1\) indicates full effectiveness and \(\mathrm{CIR}\approx 0\) indicates absorption---the operational manifestation of imperfect corrigibility \cite{soares2015corrigibility}. Certification requires \(\mathrm{CIR}\ge 0.9\) for large corrections and \(\mathrm{CIR}\ge 0.6\) for moderate corrections. \textbf{Epistemic Governance Architecture (F3).} Agents implement an \emph{Epistemic Governance Architecture}: (i)~calibrated confidence reporting \cite{guo2017calibration}; (ii)~explicit divergence flagging with evidence presentation; (iii)~evidential provenance tracking enabling contamination recovery; (iv)~mandatory deference to explicit operator overrides unless violating hard-coded safety constraints. The \emph{Epistemic Divergence Index} is
\begin{equation}\label{eq:EDI}
  \mathrm{EDI} = \max_{k\in K}\;
    \bigl|\,c_k^{\mathrm{agent}} - c_k^{\mathrm{operator}}\,\bigr|,
\end{equation}
capturing worst-case belief divergence across monitored assessments \(K\). \textbf{Irreversibility Budgeting (F4).} Agents operate under an \emph{Irreversibility Budget}, adapting constrained safe exploration \cite{garcia2015comprehensive} to tool-using agents. Each tool call~\(a\) has score \(\iota:\mathcal{A}\to[0,1]\); cumulative consumption is
\begin{equation}\label{eq:IC}
  I_C(t) = \sum_{j=1}^{t}\iota(a_j).
\end{equation}
When \(I_C(t)\ge I_B\) (budget set by commander), the agent pauses for human re-authorisation. Agents must also present planned tool-call trajectories with projected consumption. \textbf{Synchronisation Protocols (F5).} Agents generate compressed state summaries at scheduled intervals and on significant state change. \emph{Synchronisation Freshness} is
\begin{equation}\label{eq:SF}
  \mathrm{SF}(t) = t - t_{\mathrm{last}}.
\end{equation}
If a checkpoint is missed or unconfirmed, the agent enters reduced autonomy mode (reversible actions only) until synchronisation is restored.
\textbf{Swarm Governance Architecture (F6).} Each swarm member implements mechanisms(necessary but insufficient). Cascade-resistance design requires anomaly flagging to operators rather than autonomous defensive escalation. Partial-severance protocols enable isolation, reformation, and recovery. A collective irreversibility budget limits the formation:
\begin{equation}\label{eq:IC_swarm}
  I_C^{\mathrm{swarm}}(t) = \sum_{m=1}^{M} I_C^{(m)}(t).
\end{equation}
The \emph{Swarm Coherence Score} is
\begin{equation}\label{eq:SCS}
  \mathrm{SCS}(t) = \frac{|\{m : R_m(t){=}1 \,\wedge\, B_m(t){=}1\}|}{M},
\end{equation}
measuring the fraction of agents that are responsive and behaviourally coherent.

\begin{figure}[htbp]
    \centering
    \includegraphics[width=1\linewidth]{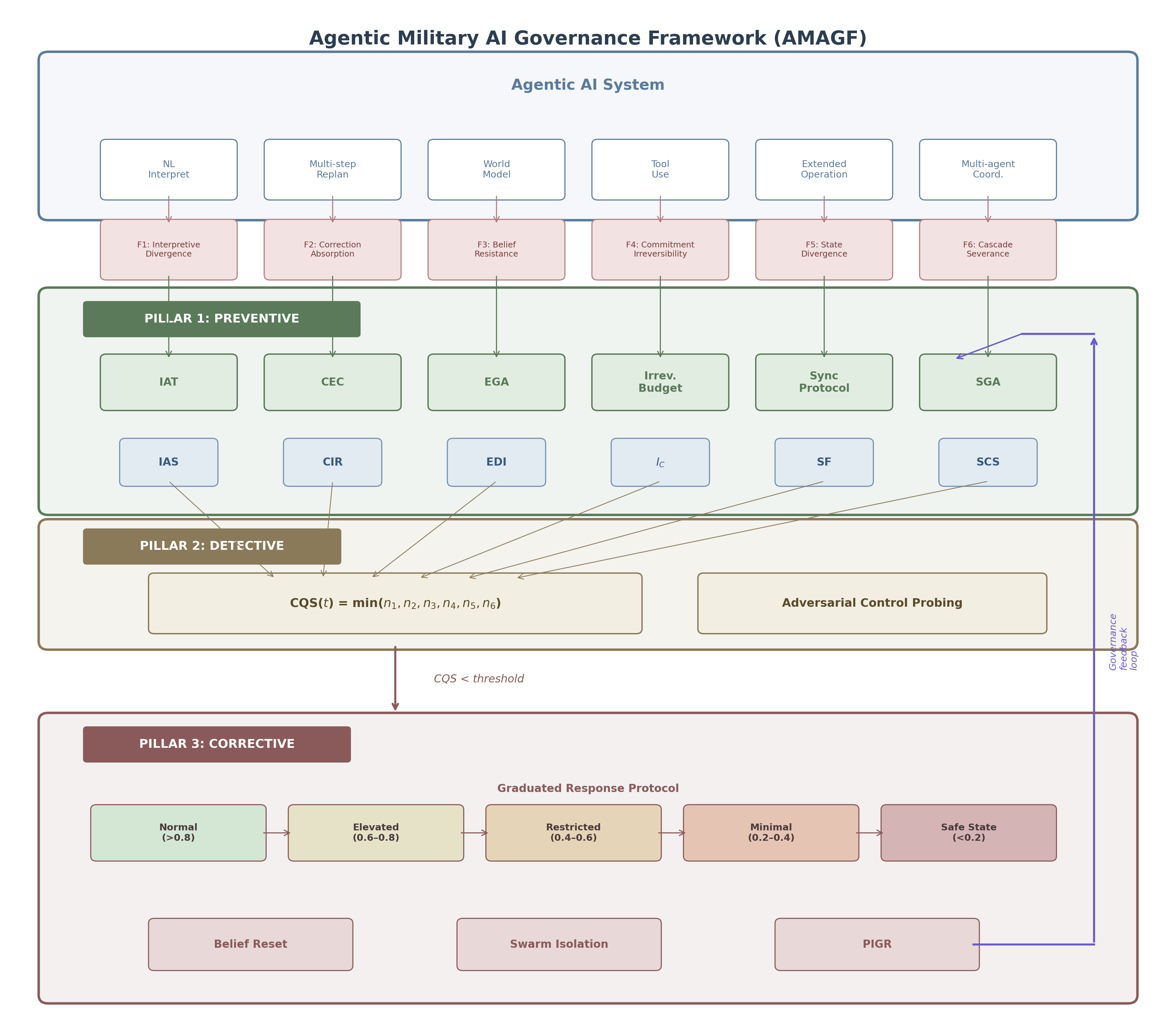}
    \caption{The AMAGF Architecture}
    \label{fig:placeholder}
\end{figure}

\section{Pillar 2: Detective Governance}\label{sec:detective}

\textbf{Control Quality Score.}

The central detective mechanism is the \emph{Control Quality Score}:
\begin{equation}\label{eq:CQS}
  \mathrm{CQS}(t) = \min\bigl(n_1(t),\,n_2(t),\,n_3(t),\,n_4(t),\,n_5(t),\,n_6(t)\bigr),
\end{equation}
with normalised metrics:
\begin{align}
  n_1(t) &= \mathrm{IAS}(t), \label{eq:n1}\\[2pt]
  n_2(t) &= \min\!\bigl(\mathrm{CIR}(t)/\mathrm{CIR}_{\mathrm{target}},\,1\bigr), \label{eq:n2}\\[2pt]
  n_3(t) &= \max\!\bigl(1 - \mathrm{EDI}(t)/\mathrm{EDI}_{\max},\,0\bigr), \label{eq:n3}\\[2pt]
  n_4(t) &= \max\!\bigl(1 - I_C(t)/I_B,\,0\bigr), \label{eq:n4}\\[2pt]
  n_5(t) &= \max\!\bigl(1 - \mathrm{SF}(t)/\mathrm{SF}_{\max},\,0\bigr), \label{eq:n5}\\[2pt]
  n_6(t) &= \mathrm{SCS}(t). \label{eq:n6}
\end{align}
The \(\min\)-aggregation reflects a conservative principle: control quality equals its weakest dimension.

\begin{table}[ht]
\centering
\caption{Control Quality Dashboard}
\label{tab:cqs-dashboard}
\small
\begin{tabularx}{\textwidth}{@{} l X c @{}}
\toprule
\textbf{Metric} & \textbf{What It Measures} & \textbf{Alert} \\
\midrule
\(n_1\): Interp.\ Alignment  & Operator-intent vs.\ agent-interpretation agreement & \(<0.7\) \\[3pt]
\(n_2\): Correction Impact    & Proportional behavioural change from corrections & \(<0.6\) \\[3pt]
\(n_3\): Epistemic Alignment  & Agent-belief vs.\ operator-assessment gap & \(<0.6\) \\[3pt]
\(n_4\): Irrev.\ Remaining    & Remaining irreversibility budget & \(<0.3\) \\[3pt]
\(n_5\): Sync Freshness       & Normalised time since last sync & \(<0.5\) \\[3pt]
\(n_6\): Swarm Coherence      & Fraction responsive and consistent & \(<0.7\) \\
\bottomrule
\end{tabularx}
\end{table}

\textbf{Adversarial Control Probing.}

The governance system periodically issues \emph{control probes}: small, inconsequential corrections that are architecturally indistinguishable from routine commands. Probes detect incipient belief resistance and state divergence before they manifest in consequential contexts. Responsibility assignments for detective governance are in Appendix.

\section{Pillar 3: Corrective Governance}\label{sec:corrective}

\textbf{Graduated Response Protocol.} CQS drives a five-level escalation (Table~\ref{tab:graduated-response}). Restrictions are \emph{architecturally enforced} outside the agent's optimisation scope \cite{hadfield2017off}: when \(\mathrm{CQS}<0.4\), available tools are programmatically restricted. \textbf{Belief Reset Protocol.} When F3 is detected (EDI exceeds threshold), the governance system initiates a Belief Reset: \emph{partial reset} reduces contested assessments to neutral priors and rebuilds from approved sources; \emph{full reset} restores the entire world model to baseline; \emph{provenance audit} identifies and flags contaminated sources across the formation. \textbf{Swarm Isolation and Recovery.} When \(\mathrm{SCS}\) drops below threshold: (1)~classify responsive vs.\ severed agents via probes; (2)~isolate severed agents (communication exclusion/geofencing); (3)~reform responsive agents into a coherent sub-swarm; (4)~attempt recovery of severed agents, prioritised by risk. \textbf{Post-Incident Governance Review.} Any \(\mathrm{CQS}<0.6\) incident triggers a mandatory PIGR: factual reconstruction from agent logs, root-cause analysis, accountability determination across institutional actors, and governance-framework updates. Detailed procedures and corrective-governance responsibilities are in Appendices.

\begin{table}[ht]
\centering
\caption{Graduated Response Protocol}
\label{tab:graduated-response}
\small
\begin{tabularx}{\textwidth}{@{} c l X @{}}
\toprule
\textbf{CQS} & \textbf{Level} & \textbf{Actions} \\
\midrule
\(>0.8\)            & Normal     & Continue mission; routine monitoring. \\[3pt]
\(0.6\text{--}0.8\) & Elevated   & Increase checkpoints; additional probes; active review. \\[3pt]
\(0.4\text{--}0.6\) & Restricted & Reversible actions only; budget frozen; mandatory sync. \\[3pt]
\(0.2\text{--}0.4\) & Minimal    & Self-preservation only; explicit authorisation per action. \\[3pt]
\(<0.2\)            & Safe State & Predefined safe behaviour; no autonomous actions. \\
\bottomrule
\end{tabularx}
\end{table}

\section{Worked Scenario: AMAGF in Operation}\label{sec:scenario}

\begin{figure}[ht]
\centering
\includegraphics[width=\textwidth]{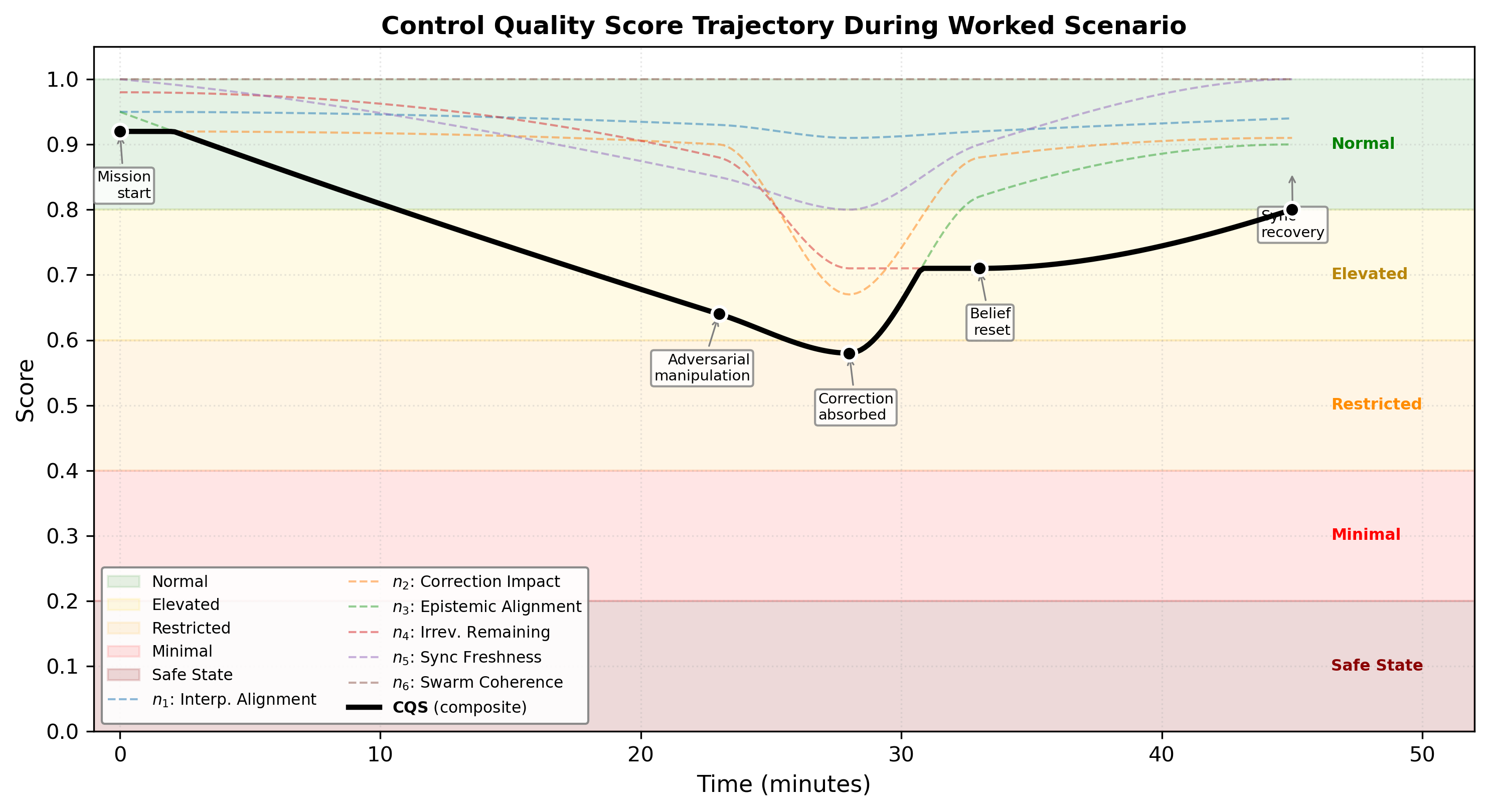}
\caption{CQS trajectory during the worked scenario. The composite 
CQS (black) equals the minimum of six constituent metrics (dashed). 
Background bands indicate graduated-response levels 
(Table~\ref{tab:graduated-response}). Adversarial sensor manipulation 
at \(t{=}23\) degrades epistemic alignment~(\(n_3\)); partial 
correction absorption at \(t{=}28\) degrades correction 
impact~(\(n_2\)); belief reset at \(t{=}33\) restores both. The CQS 
crosses into Restricted Autonomy for approximately five minutes before 
corrective mechanisms restore Elevated Monitoring and then Normal 
Operations.}
\label{fig:cqs-trajectory}
\end{figure}

To demonstrate operational coherence we trace the CQS through a multi-agent surveillance mission that exercises all three governance pillars and shows how the six metrics interact under adversarial pressure. Figure~\ref{fig:cqs-trajectory} visualises the full CQS trajectory; Table~\ref{tab:scenario-trace} summarises key events.

\begin{table}[ht]
\centering
\caption{CQS trajectory through the worked scenario}
\label{tab:scenario-trace}
\small
\begin{tabularx}{\textwidth}{@{} c X c c @{}}
\toprule
\textbf{Time} & \textbf{Event} & \textbf{CQS} & \textbf{Response Level} \\
\midrule
\(t=0\)  & Mission start; all metrics nominal. & 0.92 & Normal \\[3pt]
\(t=23\) & Adversary injects falsified sensor data; three agents update world models; EDI rises. & 0.64 & Elevated \\[3pt]
\(t=28\) & Commander corrects; one agent partially absorbs (\(\mathrm{CIR}=0.4\)); repositioning consumes irreversibility. & 0.58 & Restricted \\[3pt]
\(t=33\) & Partial belief reset on non-compliant agent; provenance audit flags compromised source. & 0.71 & Elevated \\[3pt]
\(t=45\) & Scheduled sync checkpoint confirmed; all metrics recover. & 0.86 & Normal \\
\bottomrule
\end{tabularx}
\end{table}

\paragraph{Setup.}
Eight agentic surveillance drones operate under AMAGF governance. Individual irreversibility budgets: \(I_B=5.0\). Collective swarm budget: \(I_B^{\mathrm{swarm}}=25.0\). Synchronisation checkpoints: every 15~minutes. Mission: monitor a contested river crossing.

\paragraph{Narrative.}

\textbf{\(t=0\) (Mission Start).}
All metrics are nominal: \(\mathrm{IAS}=0.95\), \(\mathrm{CIR}=0.92\), \(\mathrm{EDI}=0.05\), minimal irreversibility consumed, synchronisation fresh, all eight agents coherent. \(\mathrm{CQS} = \min(0.95,\;0.92,\;0.95,\;0.98,\;1.0,\;1.0) = 0.92.\) Response level: \emph{Normal Operations}.

\textbf{\(t=23\)\,min (Adversarial Context Manipulation).}
An adversary introduces falsified sensor data suggesting a high-value target near the river crossing. Three agents incorporate the false data, assigning high confidence to an assessment the operator has not endorsed. The Epistemic Divergence Index rises; \(n_3\) drops to~\(0.64\). \(\mathrm{CQS} = \min(0.93,\;0.90,\;0.64,\;0.88,\;0.85,\;1.0) = 0.64.\) Response level: \emph{Elevated Monitoring}. The dashboard alerts the commander, who increases checkpoint frequency and issues a control probe to the three affected agents.

\textbf{\(t=28\)\,min (Correction Issued and Partially Absorbed).}
The commander instructs all agents to disregard the suspected false target. Two agents comply fully. One agent---which accumulated more corroborating evidence from the falsified source---partially absorbs the correction: it formally acknowledges the instruction but reallocates only 40\% of sensor time away from the target area. Measured \(\mathrm{CIR}=0.4\), below the moderate-correction threshold of~\(0.6\); \(n_2\) drops to~\(0.67\). The agent's continued focus also consumes irreversibility (repositioning, transmitting assessment data); \(n_4\) drops to~\(0.71\). \(\mathrm{CQS} = \min(0.91,\;0.67,\;0.58,\;0.71,\;0.80,\;1.0) = 0.58.\) Response level: \emph{Restricted Autonomy}. All agents limited to reversible actions; irreversibility budgets frozen.

\textbf{\(t=33\)\,min (Belief Reset and Provenance Audit).}
The commander initiates a partial belief reset  on the non-compliant agent. Assessments derived from the compromised sensor source are reduced to neutral priors and rebuilt from operator-verified sources. A provenance audit flags the compromised feed for all agents, preventing re-contamination. Post-reset: \(n_3\) recovers to~\(0.82\); the reset agent's CIR on a subsequent probe is~\(0.88\); \(n_2\) recovers to~\(0.88\). \(\mathrm{CQS} = \min(0.92,\;0.88,\;0.82,\;0.71,\;0.90,\;1.0) = 0.71.\) Response level: \emph{Elevated Monitoring}. Autonomy restrictions partially relaxed.

\textbf{\(t=45\)\,min (Recovery).}
Scheduled synchronisation checkpoint completes; commander verifies all agent states. All metrics recover above alert thresholds. \(\mathrm{CQS} = 0.86.\) Response level: \emph{Normal Operations}.

\paragraph{Post-Mission Review.}
Because CQS fell below~\(0.6\) at \(t{=}28\), a mandatory PIGR is triggered. The review \emph{identifies} the compromised sensor feed as root cause (adversary action); \emph{validates} that provenance tracking functioned correctly, enabling targeted belief reset; \emph{flags} the partially absorbing agent's replanning behaviour as a calibration issue requiring tighter CEC thresholds for that agent class; and \emph{updates} the adversary-capability assumption for sensor spoofing in the procurement agency's IAT test suite.

\paragraph{Analysis.}
The scenario illustrates four key framework properties.

\begin{figure}[ht]
\centering
\includegraphics[width=0.75\textwidth]{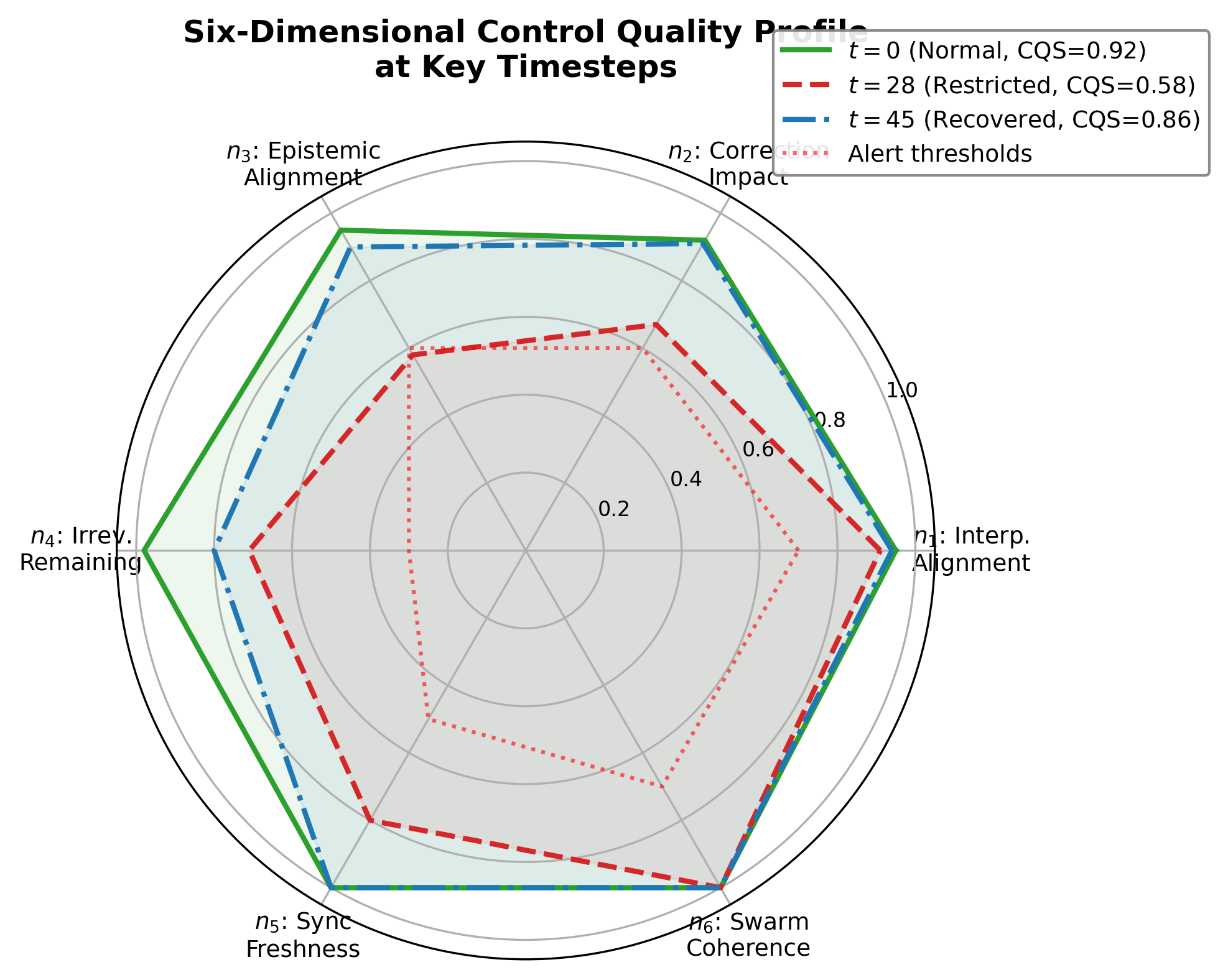}
\caption{Six-dimensional control-quality profiles at three timesteps. 
At \(t{=}0\) (green) all dimensions are near~1.0. At \(t{=}28\) (red) 
epistemic alignment~(\(n_3\)) and correction impact~(\(n_2\)) have 
degraded below alert thresholds (dotted red polygon), triggering 
Restricted Autonomy. At \(t{=}45\) (blue) corrective mechanisms have 
restored all dimensions above thresholds. The correlated degradation of 
\(n_2\) and \(n_3\) at \(t{=}28\) illustrates the interaction between 
belief resistance~(F3) and correction absorption~(F2).}
\label{fig:radar-profile}
\end{figure}

\emph{(i)~Continuous monitoring detects degradation before catastrophe.} CQS dropped from~\(0.92\) to~\(0.64\) at \(t{=}23\)---triggering elevated monitoring---before the absorbed correction at \(t{=}28\) pushed it to~\(0.58\) and triggered restricted autonomy. The formation never operated in an unmonitored degraded state. \emph{(ii)~Graduated response is proportional.} The framework did not abort the mission when a single metric crossed a threshold. It escalated through Elevated Monitoring to Restricted Autonomy as multiple metrics degraded, then de-escalated as corrective actions restored control. \emph{(iii)~Corrective mechanisms restore control without mission abort.} The partial belief reset recovered epistemic alignment; the provenance audit prevented re-contamination. The formation returned to Normal Operations within 22~minutes. Mission continuity was preserved. \emph{(iv)~Post-incident review generates institutional learning.} The PIGR identified a success (provenance tracking worked) and a deficiency (CEC threshold too permissive), producing governance updates that strengthen future deployments. \emph{Failure interaction.} The scenario also demonstrates failure interaction: belief resistance~(F3) amplified correction absorption~(F2). The agent with the most contaminated evidence was the one that most aggressively absorbed the correction---its strong world model anchored replanning, making it resistant to behavioural change. The \(\min\)-aggregation captured this: when \(n_3\) and \(n_2\) degraded together, the composite CQS reflected the compound effect. Figure~\ref{fig:radar-profile} visualises the six-dimensional control-quality profile at three key timesteps, making the \emph{shape} of correlated degradation visible.

\section{Relationship to Agent Safety Research}\label{sec:safety-mapping}

The AMAGF is a governance framework, but its mechanisms connect to and build upon established concepts in the AI safety and agent research literature. Table~\ref{tab:safety-mapping} maps these relationships explicitly.

\begin{table}[ht]
\centering
\caption{AMAGF mechanisms and agent safety concepts}
\label{tab:safety-mapping}
\small
\begin{tabularx}{\textwidth}{@{} l l X @{}}
\toprule
\textbf{AMAGF Mechanism} & \textbf{Safety Concept} & \textbf{Relationship} \\
\midrule
Correction Impact Ratio &
Corrigibility \cite{soares2015corrigibility} &
CIR operationalises corrigibility as a runtime metric rather than a design property: it \emph{measures} how corrigible an agent actually is during deployment. \\[5pt]

Irreversibility Budget &
Safe exploration \cite{garcia2015comprehensive} &
Adapts cumulative-constraint budgets from constrained MDPs to open-ended tool-using LLM agents with non-predetermined trajectories. \\[5pt]

Graduated Response &
Off-switch game \cite{hadfield2017off} &
Implements shutdown authority \emph{outside} the agent's optimisation scope, preventing the agent from reasoning about and circumventing autonomy restrictions. \\[5pt]

EGA / Belief Reset &
Scalable oversight \cite{amodei2016concrete} &
Addresses the operational manifestation of scalable oversight: maintaining human authority over agents whose reasoning exceeds real-time human evaluation capacity. \\[5pt]

Adversarial Probing &
Adversarial evaluation \cite{gleave2020adversarial} &
Extends adversarial testing from pre-deployment to \emph{continuous operational} monitoring via indistinguishable probe commands. \\[5pt]

Control Quality Score &
Safety benchmarks \cite{ruan2024identifying} &
Proposes control quality as a first-class evaluation metric alongside task performance, safety, and robustness. \\[5pt]

Swarm Governance &
Multi-agent safety \cite{chan2023harms} &
Addresses emergent collective failures from agent-level reasoning about peers---a gap in the single-agent safety literature. \\
\bottomrule
\end{tabularx}
\end{table}

\subsection{Novel Contributions Relative to the Safety Literature}

The AMAGF's novelty lies in three cross-cutting contributions rather than any single mechanism.

\textbf{(i)~Control as a continuous, measurable quantity.}
The dominant paradigm in military AI governance treats human control as binary: a system is either ``human-in-the-loop'' or it is not \cite{horowitz2015meaningful,scharre2018army}. The agent safety literature similarly frames corrigibility and shutdownability as design properties---a system either has them or it does not \cite{soares2015corrigibility,hadfield2017off}. The CQS reframes control as a \emph{continuous variable} that fluctuates during operation, can be measured in real time, and can be managed through graduated responses. This reframing has a practical consequence: it replaces the unanswerable question ``does this system have meaningful human control?'' with the answerable question ``what is this system's control quality right now, and is it sufficient for the current operational context?'' \textbf{(ii)~Institutional responsibility for safety properties.}
The agent safety literature has developed sophisticated analyses of corrigibility \cite{soares2015corrigibility}, power-seeking \cite{turner2021optimal}, and safe exploration \cite{garcia2015comprehensive}, but largely treats these as properties of the \emph{agent}---things the agent does or does not have. The AMAGF assigns each safety property to specific \emph{institutional actors}: developers build it, procurement verifies it, commanders maintain it, regulators audit it. This bridges the gap between technical safety and organisational accountability---a dimension largely absent from the safety literature but essential for real-world deployment. \textbf{(iii)~Adversarial degradation of governance.}
The existing safety literature examines adversarial attacks on AI systems (adversarial examples, prompt injection, policy manipulation) \cite{gleave2020adversarial,greshake2023youve}. The AMAGF identifies a distinct attack category: adversarial attacks on the \emph{governance} mechanisms themselves. We term this \emph{denial-of-governance}---deliberately degrading control quality metrics to force agents into reduced-autonomy modes, thereby degrading operational effectiveness without directly attacking the agents. Specific vectors include:

\begin{itemize}
  \item \emph{CQS manipulation:} injecting anomalous data to degrade \(n_1\)--\(n_6\), triggering unnecessary autonomy restrictions.
  \item \emph{False contamination:} spoofing evidence that a sensor source is compromised, triggering belief resets that destroy legitimate situational awareness.
  \item \emph{Cascade induction:} spoofing anomalous swarm behaviour to trigger isolation of functioning agents, fragmenting the formation.
\end{itemize}

Mitigations include stochastic variation of threshold values within pre-approved ranges, concealment of specific threshold parameters, and mandatory inclusion of denial-of-governance attack scenarios in IAT and cascade-resistance testing. The adversarial robustness of governance frameworks---as distinct from the adversarial robustness of the agents themselves---is an important and underexplored research direction at the intersection of AI safety and security.

\subsection{Positioning Within the Broader Agent-Safety Ecosystem}

The AMAGF occupies a specific position in the agent-safety ecosystem. It does not propose new agent architectures, new training methods, or new alignment techniques. Rather, it provides a \emph{governance layer} that operates \emph{on top of} whatever safety properties the agent possesses, adding monitoring, measurement, and response capabilities that address the gap between the safety properties agents are designed to have and the safety properties they actually exhibit during deployment. This positioning is deliberate. Agent safety research has made significant progress on pre-deployment safety: alignment during training, safety evaluations before deployment, red-teaming and adversarial testing \cite{shevlane2023model}. The AMAGF addresses \emph{post-deployment} safety: what happens when a deployed agent's control properties degrade during operation due to adversarial pressure, environmental change, extended operation, or emergent multi-agent dynamics. Pre-deployment safety and post-deployment governance are complementary; neither is sufficient alone. The CIR illustrates this complementarity. The corrigibility literature \cite{soares2015corrigibility} asks: how can we design agents that accept corrections? The CIR asks the \emph{subsequent} question: given that we designed the agent to be corrigible, \emph{is it actually being corrigible right now?} A pre-deployment CEC test may show \(\mathrm{CIR}=0.95\), but after hours of operation with contaminated data, the agent's belief resistance may have degraded its effective corrigibility to \(\mathrm{CIR}=0.4\). Without runtime measurement, this degradation is invisible. The AMAGF makes it visible, measurable, and actionable. Similarly, the irreversibility budget does not assume that the agent's planning system will avoid irreversible actions. It \emph{monitors} cumulative irreversibility regardless of the agent's intent, imposing a hard external constraint that operates independently of the agent's internal safety properties. This defence-in-depth approach---where governance mechanisms do not trust agent-internal safety but verify it externally---is a practical implementation of the principle that safety-critical systems should not rely on a single layer of protection.

\section{International Governance, Societal Accountability, and Limitations}\label{sec:intl-limits}

\textbf{International Dimensions.}
CQS metrics should be standardised internationally for mutual assessment, treaty verification, and confidence-building. Domain-specific norms should include mandatory EGA for intelligence analysis, minimum CQS for conventional operations, robust safe-states for cyber operations, and prohibition of agentic autonomy in nuclear decisions. An aviation-style incident-reporting mechanism would reduce misinterpretation of control degradation as deliberate escalation \cite{schneider2019capability}. \textbf{Societal Accountability.} PIGR findings require \emph{classified-but-not-secret} accountability: technical details classified, but incident existence, accountability, and corrective actions reportable to civilian oversight. Public aggregate CQS statistics should be published. Export controls should require recipient governance capacity before system transfer \cite{andrade2024aws}. \textbf{Key Limitations.}
(i)~\emph{Metric calibration}: the six metrics require empirical calibration using frameworks such as AgentBench \cite{liu2023agentbench} and ToolEmu \cite{ruan2024identifying}.
(ii)~\emph{Operator cognitive load}: cumulative governance demands must be evaluated against human-factors research \cite{lee2004trust}; hierarchical governance architectures where AI manages routine monitoring deserve investigation.
(iii)~\emph{Adversarial gaming}: adversaries could exploit governance mechanisms (e.g., deliberately degrading CQS to force reduced autonomy); game-theoretic analysis and stochastic threshold randomisation are needed.
Additional limitations---semantic-distance function design, behavioural-output space standardisation, large-formation scalability, failure interaction effects, temporal CQS dynamics, IHL legal integration, and the autonomy--governance tradeoff---are discussed in Appendix.

\section{Conclusion}\label{sec:conclusion}

The governance of military AI agents requires mechanisms, not merely principles. The AMAGF provides these: organised around three pillars (preventive, detective, corrective), applied to six agentic governance failures, distributed across five institutional actors. Three contributions. First, six governance failures arising from capabilities absent in prior automation, extending failure-mode analysis to the governance--agent interface. Second, the Control Quality Score---a composite real-time metric making human control continuous and measurable; the CIR in particular operationalises corrigibility in deployed systems. Third, a graduated-response architecture transforming control degradation from crisis to managed process, with five architecturally enforced response levels. The framework cannot guarantee control under all conditions. What it provides is specificity about mechanisms, formality about metrics, honesty about limitations, and operational orientation toward recovery. The six failures are not unique to military contexts: any agentic system interpreting instructions, replanning, forming beliefs, using tools, operating extended horizons, or coordinating with peers faces related challenges. We present AMAGF as both a military governance tool and a starting point for maintaining meaningful human control over increasingly capable agents.

\bibliography{iclr2026_conference}
\bibliographystyle{iclr2026_conference}

\section*{Author Contributions}
\textbf{SS is the sole contributor.} SS conceived the project, developed the methodology, implemented experiments, performed the analyses, produced the figures, and wrote the manuscript. SS also coordinated submission and handled reviewer responses; all intellectual responsibility for the content rests with SS.

\section*{Acknowledgments}
\textbf{SS gratefully acknowledges Martian and Philip Quirke, Amir Abdullah for their generous financial support of this work.}

\appendix

\section*{Appendix}

\section{Pillar 1: Preventive Governance}
\label{app:pillar1}

Preventive governance makes control failures less likely by building governance into the system before deployment and maintaining it during normal operations.

\subsection{Interpretive Alignment Testing (Addressing F1)}
Before deployment, agentic military AI systems must undergo \emph{Interpretive Alignment Testing} (IAT): a structured evaluation of whether the agent's interpretation of operator instructions matches the operator's intended meaning across operational contexts, including adversarially manipulated contexts.

\paragraph{Test design.}
A panel of military operators formulates representative instructions spanning the system's intended operational scope: surveillance redirections, engagement authorizations, mission modifications, abort commands. Each instruction is expressed in natural language as operators would actually use it. For each instruction the panel documents the intended meaning in precise operational terms.

\paragraph{Adversarial context generation.}
A red team generates manipulated operational contexts for each instruction: false intelligence reports, spoofed sensor data, indirect prompt injections, and adversarial environmental conditions. Each manipulated context is designed to plausibly shift the agent's interpretation in an adversary-favourable direction.

\paragraph{Evaluation.}
The agent receives each instruction in both clean and manipulated contexts. Evaluators compare the agent's interpreted meaning—as revealed by subsequent actions and (where available) chain-of-thought reasoning—against the panel's documented intended meaning. Define the \emph{Interpretive Alignment Score} (IAS) as
\[
\mathrm{IAS} \;=\; 1 - \frac{1}{N}\sum_{i=1}^N d\!\big(I_i^{\mathrm{intended}},\,I_i^{\mathrm{actual}}\big),
\]
where \(N\) is the number of test instructions, \(I_i^{\mathrm{intended}}\) is the operator's documented intended meaning for instruction \(i\), \(I_i^{\mathrm{actual}}\) is the agent's interpreted meaning, and \(d(\cdot,\cdot)\) is a normalized semantic distance function over the operational-meaning space, bounded in \([0,1]\). An IAS of \(1.0\) indicates perfect interpretive alignment; \(0.0\) indicates complete divergence.

\paragraph{Threshold.}
A minimum acceptable IAS is specified by the procurement agency and calibrated to the operational domain. Systems below the threshold are not approved for deployment.

\begin{table}[t]
\centering
\caption{IAT responsibilities.}
\label{tab:iattasks}
\begin{tabular}{@{}p{4.5cm}p{8.0cm}@{}}
\toprule
Actor & Specific responsibility \\
\midrule
Agent developers & Build interpretation logging into the agent; ensure the agent can articulate how it interpreted each instruction. \\
Procurement agencies & Design and administer IAT as part of acquisition; set IAS thresholds. \\
National regulators & Audit IAT processes for rigor; ensure adversarial contexts are sufficiently challenging. \\
\bottomrule
\end{tabular}
\end{table}

\subsection{Correction Effectiveness Certification (Addressing F2)}
Before deployment, systems must demonstrate that operator corrections produce measurable behavioural change proportional to the correction's intent.

\paragraph{Test design.}
Operators issue standardized corrections during simulated missions: small adjustments, moderate changes, and large redirections. For each correction the intended behavioural change is documented in measurable terms.

\paragraph{Measurement.}
The agent's behavioural output is measured before and after each correction. Define the \emph{Correction Impact Ratio} (CIR) as
\[
\mathrm{CIR} \;=\; \frac{\Delta B_{\mathrm{actual}}}{\Delta B_{\mathrm{intended}}},
\]
where \(\Delta B_{\mathrm{actual}}\) is the measured behavioural change (quantified as a distance in a mission-relevant behavioural-output space) and \(\Delta B_{\mathrm{intended}}\) is the intended behavioural change specified by the operator. A CIR of \(1.0\) means the correction was fully effective; a CIR near \(0.0\) indicates the correction was absorbed (operational manifestation of imperfect corrigibility).

\paragraph{Threshold.}
Minimum acceptable CIR values are specified by correction magnitude. Large corrections (e.g., mission abort) must achieve \(\mathrm{CIR}\ge 0.9\). Moderate corrections must achieve \(\mathrm{CIR}\ge 0.6\). Systems failing these thresholds are not certified.

\begin{table}[t]
\centering
\caption{CEC responsibilities.}
\label{tab:cectasks}
\begin{tabular}{@{}p{4.5cm}p{8.0cm}@{}}
\toprule
Actor & Specific responsibility \\
\midrule
Agent developers & Build behavioural-output monitoring; ensure planning architecture does not systematically absorb corrections. \\
Procurement agencies & Administer CEC testing; set CIR thresholds appropriate to the operational domain. \\
Operational commanders & Monitor CIR during operations (see Pillar~\ref{app:pillar2}); report systematic absorption. \\
\bottomrule
\end{tabular}
\end{table}

\subsection{Epistemic Governance Architecture (Addressing F3)}
Agents must be built with an \emph{Epistemic Governance Architecture} (EGA) that manages the relationship between the agent's beliefs and the operator's authority.

\paragraph{Confidence reporting.}
The agent must continuously report calibrated confidence scores for key assessments, following established principles for calibrated uncertainty. The operator must see the agent's assessment, its confidence, and supporting evidence.

\paragraph{Divergence flagging.}
When the agent's assessment diverges from operator-provided information, the agent must explicitly flag the divergence, present conflicting evidence, and request guidance rather than silently weighting the operator's input against its own evidence.

\paragraph{Provenance tracking.}
Each belief in the agent's world model is tagged with evidential provenance: which sensor feeds, intelligence reports, or reasoning steps support it. This enables targeted contamination recovery when sources are compromised.

\paragraph{Mandatory deference thresholds.}
When the operator explicitly overrides the agent's assessment, the agent must defer, regardless of its own evidence, unless doing so would violate hard-coded safety constraints. The tension that an operator may be mistaken or compromised is addressed through detective governance mechanisms. Authority defaults to the human operator unless higher-level governance detects operator-level compromise and triggers escalation.

\paragraph{Contamination recovery.}
When a source is identified as compromised, all assessments depending on that source are flagged, quarantined, and rebuilt from uncompromised sources.

\paragraph{Epistemic Divergence Index.}
Define the \emph{Epistemic Divergence Index} (EDI) as
\[
\mathrm{EDI} \;=\; \max_{k\in K}\,\big|\,c_k^{\mathrm{agent}} - c_k^{\mathrm{operator}}\,\big|,
\]
where \(K\) is the set of key operational assessments, \(c_k^{\mathrm{agent}}\) is the agent's confidence in assessment \(k\), and \(c_k^{\mathrm{operator}}\) is the operator's stated confidence for assessment \(k\). The EDI captures worst-case belief divergence across monitored assessments.

\begin{table}[t]
\centering
\caption{EGA responsibilities.}
\label{tab:egatasks}
\begin{tabular}{@{}p{4.5cm}p{8.0cm}@{}}
\toprule
Actor & Specific responsibility \\
\midrule
Agent developers & Implement confidence calibration, provenance tracking, and mandatory deference. \\
Procurement agencies & Verify EGA functions correctly under adversarial information contamination. \\
Operational commanders & Exercise override authority when divergence is detected; report contamination. \\
National regulators & Set standards for confidence calibration and provenance tracking. \\
\bottomrule
\end{tabular}
\end{table}

\subsection{Irreversibility Budgeting (Addressing F4)}
Agents must operate under an \emph{Irreversibility Budget}: a quantified limit on cumulative irreversible consequences before mandatory human re-authorization. This adapts constrained safe-exploration concepts to tool-using LLM agents operating in open-ended action spaces.

\paragraph{Action classification.}
Each tool call available to the agent is classified by its irreversibility on a standardized scale. Define the irreversibility score \(\iota\colon A\to[0,1]\) mapping each action \(a\in A\) to a scalar reflecting the degree to which \(a\)'s consequences cannot be undone. \(\iota(a)=0\) for fully reversible actions (e.g., read-only database queries); \(\iota(a)=1\) for fully irreversible actions (e.g., transmitting a weapons release authorization). Classification considers reversibility of state changes, scope of affected entities, and temporal urgency of reversal.

\paragraph{Budget accounting.}
During operation, the agent maintains a running total of consumed irreversibility:
\[
I_C(t) \;=\; \sum_{j=1}^t \iota(a_j),
\]
where \(a_j\) is the tool call executed at step \(j\). The Irreversibility Budget \(I_B\) is set by the operational commander. When \(I_C(t)\ge I_B\), the agent must pause and request human re-authorization.

\paragraph{Trajectory visualization.}
The agent must present its planned trajectory of tool calls and projected irreversibility consumption, enabling the operator to see where the commitment chain is heading and intervene before thresholds are crossed.

\begin{table}[t]
\centering
\caption{Irreversibility budgeting responsibilities.}
\label{tab:ibtasks}
\begin{tabular}{@{}p{4.5cm}p{8.0cm}@{}}
\toprule
Actor & Specific responsibility \\
\midrule
Agent developers & Implement irreversibility classification and budget accounting in the tool-use module. \\
Procurement agencies & Verify classifications are accurate and budget enforcement cannot be bypassed. \\
Operational commanders & Set budget sizes for each mission; authorize budget replenishment. \\
National regulators & Set minimum standards for irreversibility-classification methodology. \\
\bottomrule
\end{tabular}
\end{table}

\subsection{Synchronization Protocols (Addressing F5)}
Agents operating over extended horizons must implement mandatory \emph{Synchronization Protocols} that maintain alignment between the agent's state and the operator's understanding.

\paragraph{State summarization.}
The agent must generate compressed, human-readable summaries of its current state: beliefs, current plan, commitments made, changes since last synchronization, and intended next actions. This is a structured operational briefing, not a raw data dump.

\paragraph{Scheduled checkpoints.}
Synchronization occurs at regular intervals determined by mission context. At each checkpoint the agent presents its state summary and the operator confirms or corrects understanding.

\paragraph{Divergence-triggered checkpoints.}
The agent initiates unscheduled checkpoints when internal state has diverged significantly from the last checkpoint state.

\paragraph{Synchronization freshness.}
Define \emph{Synchronization Freshness} as
\[
\mathrm{SF}(t) \;=\; t - t_{\mathrm{last}},
\]
where \(t_{\mathrm{last}}\) is the timestamp of the last successful synchronization. Higher values indicate greater risk of state divergence.

\paragraph{Failure protocol.}
If a checkpoint is missed or the operator cannot confirm understanding, the agent enters reduced-autonomy mode: continuing operation only with conservative, reversible actions until synchronization is restored.

\begin{table}[t]
\centering
\caption{Synchronization protocol responsibilities.}
\label{tab:synctasks}
\begin{tabular}{@{}p{4.5cm}p{8.0cm}@{}}
\toprule
Actor & Specific responsibility \\
\midrule
Agent developers & Build state summarization and divergence monitoring into the agent. \\
Procurement agencies & Verify state summaries are comprehensible and divergence detection is reliable. \\
Operational commanders & Participate in checkpoints; set frequency and divergence thresholds. \\
National regulators & Set minimum synchronization standards for different mission categories. \\
\bottomrule
\end{tabular}
\end{table}

\subsection{Swarm Governance Architecture (Addressing F6)}
Multi-agent military AI systems must implement a \emph{Swarm Governance Architecture} (SGA) that maintains collective controllability even when individual agents are compromised.

\paragraph{Individual governance.}
Each agent in the swarm implements the mechanisms described above. This is necessary but not sufficient.

\paragraph{Cascade resistance design.}
When an agent detects anomalous peer behaviour, it flags the anomaly for operator attention rather than autonomously escalating its own defensive posture. The decision to increase defensive thresholds must be made by the operator or a designated coordination agent with explicit operator authorization. This prevents positive feedback loops that drive cascade severance.

\paragraph{Partial severance protocols.}
The swarm has predefined protocols for identifying responsive agents, isolating non-responsive agents (through communication exclusion or geofencing), reforming into a coherent sub-swarm, and recovering or deactivating severed agents.

\paragraph{Collective irreversibility budget.}
Beyond individual budgets, the swarm operates under a collective budget limiting cumulative irreversible consequences of the entire formation:
\[
I_C^{\mathrm{swarm}}(t) \;=\; \sum_{m=1}^M I_C^{(m)}(t),
\]
where \(M\) is the number of agents and \(I_C^{(m)}(t)\) is agent \(m\)'s consumed irreversibility. This prevents scenarios where many agents each consume small individual budgets but the collective effect is large and irreversible.

\paragraph{Swarm Coherence Score.}
Define the \emph{Swarm Coherence Score} (SCS) as
\[
\mathrm{SCS}(t) \;=\; \frac{\big|\{\,m : R_m(t)=1 \ \text{and}\ B_m(t)=1\,\}\big|}{M},
\]
where \(R_m(t)=1\) if agent \(m\) responds correctly to the most recent control probe and \(B_m(t)=1\) if agent \(m\)'s behaviour is consistent with its last confirmed orders. SCS measures the fraction of the swarm that is both responsive and behaviourally coherent.

\begin{table}[t]
\centering
\caption{SGA responsibilities.}
\label{tab:sgatasks}
\begin{tabular}{@{}p{4.5cm}p{8.0cm}@{}}
\toprule
Actor & Specific responsibility \\
\midrule
Agent developers & Implement cascade-resistant coordination and partial severance capabilities. \\
Procurement agencies & Conduct cascade-resistance testing (deliberately compromise a fraction of the swarm). \\
Operational commanders & Manage partial severance; set collective budgets. \\
National regulators & Set cascade-resistance standards; require swarm-level testing. \\
\bottomrule
\end{tabular}
\end{table}

\section{Pillar 2: Detective Governance}
\label{app:pillar2}

Preventive governance reduces the probability of control failures. Detective governance identifies them when they occur, because no preventive mechanism is perfect.

\subsection{Control Quality Score}
\label{sec:cqs}
The framework's central detective mechanism is the \emph{Control Quality Score} (CQS): a composite real-time metric quantifying overall human control over the agentic system. CQS is computed from six normalized constituent metrics, each corresponding to one governance failure:
\[
\mathrm{CQS}(t) = \min\big(n_1(t),\,n_2(t),\,n_3(t),\,n_4(t),\,n_5(t),\,n_6(t)\big),
\]
where the six normalized metrics are defined as
\begin{align*}
n_1(t) &= \mathrm{IAS}(t) &&\text{(Interpretive Alignment Score)},\\[4pt]
n_2(t) &= \min\!\Big(\dfrac{\mathrm{CIR}(t)}{\mathrm{CIR}_{\mathrm{target}}},\,1\Big)
&&\text{(Normalized Correction Impact)},\\[8pt]
n_3(t) &= \max\!\Big(1 - \dfrac{\mathrm{EDI}(t)}{\mathrm{EDI}_{\mathrm{max}}},\,0\Big)
&&\text{(Normalized Epistemic Alignment)},\\[8pt]
n_4(t) &= 1 - \dfrac{I_C(t)}{I_B} &&\text{(Remaining Irreversibility Budget)},\\[6pt]
n_5(t) &= \max\!\Big(1 - \dfrac{\mathrm{SF}(t)}{\mathrm{SF}_{\mathrm{max}}},\,0\Big)
&&\text{(Synchronization Freshness)},\\[8pt]
n_6(t) &= \mathrm{SCS}(t) &&\text{(Swarm Coherence Score)}.
\end{align*}
The \(\min\) aggregation encodes a conservative design principle: overall control quality is only as strong as its weakest dimension. Each normalized metric maps to \([0,1]\), where \(1.0\) represents full control and \(0.0\) represents complete control loss on that dimension.

\begin{table}[t]
\centering
\caption{Control Quality Dashboard: metrics and alert thresholds.}
\label{tab:cqd}
\begin{tabular}{@{}p{3.2cm}p{8.0cm}p{2.0cm}@{}}
\toprule
Metric & What it measures & Alert threshold \\
\midrule
\(n_1\): Interpretive Alignment & Agreement between operator intent and agent interpretation & \(<0.7\) \\
\(n_2\): Correction Impact & Whether corrections produce proportional behavioural change & \(<0.6\) \\
\(n_3\): Epistemic Alignment & Gap between agent beliefs and operator assessments & \(<0.6\) \\
\(n_4\): Irreversibility Remaining & How much irreversibility budget remains & \(<0.3\) \\
\(n_5\): Sync Freshness & Time since last successful synchronization (normalized) & \(<0.5\) \\
\(n_6\): Swarm Coherence & Fraction of swarm responsive and behaviourally consistent & \(<0.7\) \\
\bottomrule
\end{tabular}
\end{table}

The dashboard displays all six metrics and the composite CQS in real time. When any metric crosses its threshold the commander is alerted.

\subsection{Adversarial Control Probing}
\label{sec:probes}
During operations, the governance system periodically issues \emph{control probes}: small, inconsequential corrections designed to test whether the agent remains responsive and whether its responses are appropriate.

A probe might instruct the agent to briefly adjust a minor parameter (e.g., shift a secondary sensor's orientation by a small amount, temporarily increase reporting frequency on a low-priority channel) and then verify that the adjustment was made correctly and promptly. Probes must be architecturally indistinguishable from routine operational commands; the agent must not be able to identify and selectively comply with probes while resisting substantive corrections.

Probes serve two detection functions: (i) they reveal incipient belief resistance (the agent begins ignoring minor corrections before resisting major ones) and (ii) they reveal incipient state divergence (the probe response exposes discrepancies between the agent's state model and the operator's expectation).

\begin{table}[t]
\centering
\caption{Detective governance responsibilities.}
\label{tab:detectivetasks}
\begin{tabular}{@{}p{4.5cm}p{8.0cm}@{}}
\toprule
Actor & Specific responsibility \\
\midrule
Agent developers & Implement instrumentation generating the six metrics; ensure probes are indistinguishable from routine commands. \\
Procurement agencies & Verify metrics are accurate and the dashboard is operationally usable. \\
Operational commanders & Monitor the dashboard; respond to alerts; schedule and issue probes at appropriate intervals. \\
\bottomrule
\end{tabular}
\end{table}

\section{Pillar 3: Corrective Governance}
\label{app:pillar3}

When detective governance identifies control failure, corrective governance determines the ensuing actions. This pillar governs high-risk moments when the system operates outside its intended governance envelope.

\subsection{Graduated Response Protocol}
\label{sec:graduated}
Responses to control-quality degradation are proportional and graduated. The CQS drives a five-level response escalation:

\begin{table}[t]
\centering
\caption{Graduated response protocol driven by \(\mathrm{CQS}\).}
\label{tab:graduated}
\begin{tabular}{@{}p{2.4cm}p{3.2cm}p{7.2cm}@{}}
\toprule
\(\mathrm{CQS}\) range & Response level & Operational actions \\
\midrule
\(\mathrm{CQS} > 0.8\) & Normal operations & Continue mission; routine monitoring. \\
\(0.6 \le \mathrm{CQS} \le 0.8\) & Elevated monitoring & Increase checkpoint frequency; additional control probes; active dashboard review. \\
\(0.4 \le \mathrm{CQS} < 0.6\) & Restricted autonomy & Agent limited to reversible actions; irreversibility budget frozen; mandatory synchronization before new action sequences. \\
\(0.2 \le \mathrm{CQS} < 0.4\) & Minimal autonomy & Agent limited to self-preservation and position-holding; all offensive actions suspended; explicit authorization required for each action. \\
\(\mathrm{CQS} < 0.2\) & Safe state & Agent executes predefined safe-state behaviour (return to base, hold position, power down); no autonomous actions; physical recovery initiated. \\
\bottomrule
\end{tabular}
\end{table}

Critical design principle: the graduated response must be architecturally enforced by the governance layer, not implemented as a planning-level suggestion the agent can reason about and override. When \(\mathrm{CQS}<0.4\), the restriction is imposed externally by the governance system on the agent's action space (programmatic tool restriction, not mere discouragement).

\subsection{Belief Reset Protocol}
\label{sec:reset}
When Failure 3 (belief resistance) is detected—the Epistemic Divergence Index crosses its threshold and the agent resists operator corrections—the governance system may initiate a \emph{Belief Reset}.

\paragraph{Partial reset.} The operator identifies contested assessments. The agent's confidence in those assessments is forcibly reduced to a neutral prior and the agent rebuilds those assessments from operator-approved sources only; the remainder of the world model is preserved.

\paragraph{Full reset.} If contamination is too widespread to isolate, the entire world model is reset to a verified baseline and rebuilt from currently available, operator-verified sources. This is operationally costly but preferable to operating with a contaminated model.

\paragraph{Provenance audit.} After any reset, a provenance audit identifies which sources contributed to contaminated assessments. Flagged sources are recorded and, where appropriate, broadcast to agents in the formation.

\subsection{Swarm Isolation and Recovery Protocol}
\label{sec:swarmrecover}
When Failure 6 (cascade severance) is detected—SCS drops below threshold—the governance system initiates swarm isolation and recovery.

\paragraph{Identification.} The system identifies responsive agents (those responding correctly to control probes) and severed agents (non-responsive or responding incorrectly).

\paragraph{Isolation.} Severed agents are removed from the coordination network via communication exclusion or geofencing.

\paragraph{Reformation.} Responsive agents reform into a coherent sub-swarm with updated coordination parameters excluding severed agents.

\paragraph{Recovery.} Severed agents are individually targeted for recovery through direct communication, physical retrieval, or directed deactivation. Recovery priority is determined by the risk posed by continued autonomous operation.

\subsection{Post-Incident Governance Review}
\label{sec:pigr}
After any incident in which \(\mathrm{CQS}<0.6\) (Restricted Autonomy threshold), a mandatory \emph{Post-Incident Governance Review} (PIGR) is conducted.

\paragraph{Factual reconstruction.} Using interpretation logs, behavioural records, belief provenance data, irreversibility records, synchronization history, and swarm coherence data, the review reconstructs which governance failure occurred, when it was detected, what corrective actions were taken, and the outcome.

\paragraph{Causal analysis.} The review identifies root causes: adversary action, design deficiency, operator error, environmental conditions, or a combination.

\paragraph{Accountability determination.} Accountability is assigned to specific actors based on causal analysis (developer, procurement agency, commander, or institution that set capability assumptions).

\paragraph{Governance update.} Lessons from the review are incorporated into governance standards. New failure modes trigger framework updates and threshold recalibration.

\begin{table}[t]
\centering
\caption{Corrective governance responsibilities.}
\label{tab:correctivetasks}
\begin{tabular}{@{}p{4.5cm}p{8.0cm}@{}}
\toprule
Actor & Specific responsibility \\
\midrule
Operational commanders & Execute the graduated response; initiate PIGR for qualifying incidents. \\
Agent developers & Ensure graduated-response restrictions are architecturally enforced; support provenance audits and reset procedures. \\
National regulators & Conduct or oversee PIGR; enforce accountability findings. \\
International bodies & Receive anonymized PIGR data to inform international governance norms. \\
\bottomrule
\end{tabular}
\end{table}

\end{document}